\documentclass[a4paper]{jpconf}
\usepackage{graphicx}
\usepackage{graphics}
\usepackage{amssymb}
\usepackage{amsmath}
\usepackage{array}
\usepackage{bm}
\usepackage[normalem]{ulem}

\def\refe@jnl#1{{#1}}
\def\aj{\refe@jnl{Astron.~J.}}                  
\def\araa{\refe@jnl{Annu.~Rev.~Astron.~Astrophys.}}
\def\apj{\refe@jnl{Astrophys.~J.}}                 
\def\apjl{\refe@jnl{Astrophys.~J.~Lett.}} 
\def\apjs{\refe@jnl{Astrophys.~J.~Suppl.}}         
\def\aap{\refe@jnl{Astron.~Astrophys.}}            
\def\mnras{\refe@jnl{Mon.~Not.~R.~Astron.~Soc.}}   
\def\prd{\refe@jnl{Phys.~Rev.~D}}        
\def\fcp{\refe@jnl{Fund.~Cos.~Phys.}}  
\def\physrep{\refe@jnl{Phys.~Rep.}}   
\def\physlett{\refe@jnl{Phys.~Lett.}}
\def\nat{\refe@jnl{Nat.}}

\def\fun#1#2{\lower3.6pt\vbox{\baselineskip0pt\lineskip.9pt
\ialign{$\mathsurround=0pt#1\hfill##\hfil$\crcr#2\crcr\sim\crcr}}}

\def\rn{} 
\def\nnn#1#2 #3{#2. #3. #1}                    
\def\nnnn#1 #2 #3 #4{#2. #3. #4. #1}            
\def\nnnnn#1 #2 #3 #4 #5{#2. #3. #4 #5. #1}     
  
\def\rf#1;#2;#3;#4;#5 {{\frenchspacing\par\rn#1, #3 {\bf #4}, #5 (#2). \par}} 
\def\rfbook#1;#2;#3;#4;#5 {{\frenchspacing\par\rn#1, {\it #3} (#5, #4,#2).\par}} 
\def\rfprep#1;#2;#3 {{\par\frenchspacing\rn#1, #3 (#2).\par}}

\newcommand{\nc}{\newcommand}

\nc{\beq}[1]{\begin{equation}\label{#1}} \nc{\eeq}{\end{equation}}

\nc{\inv}[1]{\frac{1}{#1}}

\def\gsim{\; \raise0.3ex\hbox{$>$\kern-0.75em
\raise-1.1ex\hbox{$\sim$}}\; }

\nc{\hn}{\hat{\textbf{n}}}
\nc{\bfi}[1]{\textit{#1}}

\begin{document}


\title{Cosmic Microwave and Infrared Backgrounds cross-correlation for ISW detection}


\author[Ili\'c]{St\'ephane Ili\'c}

\address{Universit\'e Paris-Sud, Institut d'Astrophysique Spatiale, UMR8617, Orsay, F-91405 \\ \& CNRS, Orsay, F-91405}

\ead{stephane.ilic@ias.u-psud.fr}



\begin{abstract} 
We investigate the cross-correlation between the cosmic infrared and microwave backgrounds (CIB \& CMB) anisotropies through the integrated Sachs-Wolfe effect.~We first describe the CIB anisotropies using a linearly biased power spectrum, then derive the theoretical angular power spectrum of the CMB-CIB cross-correlation for different instruments and frequencies.
We discuss the detectability of the ISW signal by performing a signal-to-noise (SNR) analysis with our predicted spectra.  
The significances obtained range from $6\sigma$ to $7\sigma$ in an ideal case, depending on the frequency~; in realistic cases which account for the presence of noise including astrophysical contaminants, the results span the range $2$-$5\sigma$, depending strongly on the major contribution to the noise term.
\end{abstract} 





\section{Introduction}
\label{sec:introduction}

The discovery of the acceleration of the expansion of the Universe, made through supernov{\ae} observations at the end of the last century, led us to hypothesize the existence of an unknown ``Dark Energy", contributing 70\% to the energy budget of our Universe.
In this presentation, we focus on a specific probe of the Dark Energy, namely the integrated Sachs-Wolfe (ISW) effect~: This effect of gravitational origin induces secondary anisotropies in the CMB and is due to large-scale structures. The gravitional potentials of the latter are slowly decaying in a $\Lambda$-dominated Universe, and therefore give a net difference in energy to the CMB photons that travel across them.
This effect shows in the power spectrum of the Cosmic Microwave Background (CMB) temperature anisotropies at large angular scales 
but its small amplitude and the cosmic variance at those $\ell$ make its direct detection very challenging, if not impossible, when using only the CMB itself. To circumvent this limitation, cosmologists have devised a way to exploit the link between this imprint on the CMB and the large-scale structures causing it, by simply cross-correlating the CMB with matter density maps (galaxy maps in practice) and then comparing the results to a null hypothesis and to what is expected from theory.

During the last decade or so, a growing interest has risen in this field thanks to the development of large galaxy surveys in many wavelengths. However, this method has yet to produce a definitive and conclusive detection of the ISW effect from current surveys, with significances ranging so far from negligible up to more than 4$\sigma$, and with sometimes conflicting results
throughout the literature. We make here a short report of our work presented in Ili\'c et al., 2011~(\cite{my_paper}). The originality of our work is to consider the Cosmic Infrared Background (CIB), first discovered by Puget et al. in 1996. This background, visible roughly from $10$ to $1000~\mu m$ in wavelength, arises from the accumulated emission of star-forming galaxies spanning a large range of redshifts including our current Dark Energy dominated era. 
The CIB also features anisotropies that are underlined by the galaxy density field and thus the matter density fluctuations. It is therefore reasonable to expect that it has a positive correlation with the CMB through the aforementioned ISW effect.


\section{Modelling the expected signal}
\label{sec:cib_isw}


\subsection{CIB anisotropies}
\label{sec:cib_anis}

Ever since its discovery, many efforts have been deployed to detect the CIB and its anisotropies with increasing precision. The most recent studies on the subject make use of the halo occupation distribution and the Dark Matter halos properties, in order to predict the power spectrum of these anisotropies. Such models are particularly useful when describing the small, non-linear scales of the CIB. Since we focus here on the ISW effect which only concerns much larger scales, we can use a simpler model for the CIB anisotropies power spectrum, similar to the description made by Knox et al. in \cite{Knox01}. We first write the CIB temperature anisotropies as the following line-of-sight integral~:
\beq{dtcib}
\delta T_{\rm{CIB}}(\hn,\nu)=\int_{\eta_{\rm{far}}}^{\eta_0} dz \, \frac{d\eta}{dz} \, a(z) \, \delta j((\eta_0-\eta)\hn,\nu,z) \ 
\eeq
with $\delta j$ being the emissivity fluctuations of the CIB. 
Again, similarly to \cite{Knox01}, we assume that the CIB anisotropies are direct tracers of the matter density fluctuations up to a factor $b_j(\nu,z)$, a frequency- and redshift-dependent matter-emissivity bias : ${\delta j((\eta_0-\eta)\hn,\nu,z)}/{\bar{j}(\nu,z)} \! = \! b_j(\nu,z) \delta((\eta_0-\eta)\hn,z)$ where
$\bar{j}(\nu,z)$ is the mean emissivity per comoving unit volume at frequency $\nu$ as a function of redshift $z$, and is derived using the empirical, parametric model of \cite{Beth}. We describe the matter density field $\delta$ in our analysis by a linear power spectrum valid for the scales of interest in our work ($\ell<100$, comprising most of the ISW signal). Lastly we choose the linear bias~
$b_j(\nu,z)$ to be constant in redshift~:  $b_j(\nu,z) \! = \! b_{\, \rm{lin}}(\nu)$. As detailed in Ili\'c et al. (\cite{my_paper}), to obtain it at each frequency, we compute the value of $b_{\, \rm{lin}}$ that gives the best agreement between our predicted linear CIB power spectrum and the one obtained from the \textit{Planck} data and published in \cite{CIB_Planck}. We get the following values for these bias factors~: $b_{\, \rm{lin}}=1.74$ at 857~GHz, $2.09$ at 545~GHz, $2.63$ at 353~GHz and $2.51$ at 217~GHz ; these values are consistent with the results found in the literature. The linear bias we obtain this way increases with the wavelength~: this is coherent with the fact that as we go deeper into the infrared, the galaxies probed are more luminous at higher $z$. They reside in more massive and rarer halos, and are therefore more biased.


\subsection{Correlation with the ISW}

In the CMB anisotropies, the temperature contribution due to the ISW effect is an integral over the conformal time of the growth rate of the gravitational potentials~:
\beq{dtisw}
\delta T_{\rm{ISW}}(\hn)=\int_{\eta_r}^{\eta_0} d\eta \ e^{-\tau(\eta)} \ (\dot{\Phi}-\dot{\Psi})[(\eta_0-\eta)\hn,\eta]
\eeq
where $\eta_r$ is some initial time deep in the radiation era, $\Phi$ and $\Psi$ are the Newtonian gauge gravitational potentials,
$\tau(\eta)$ is the optical depth, and the dot denotes differentiation with respect to $\eta$. Using this equation and Eq.~(\ref{dtcib}), and going to Fourier space, we can express the CMB-CIB cross-power spectrum at a frequency $\nu$~:
\begin{multline}
\label{crosscl}
C_{\ell}^{\rm{cr}}(\nu) = 4\pi \frac{9}{25} \int \! \frac{dk}{k} \Delta_{\mathcal{R}}^2 \times \int_{\eta_0}^{\eta_r} \! d\eta \ e^{-\tau(\eta)} \ j_{\ell}(k[\eta-\eta_0]) \ (c_{\Psi\Phi}\dot{\psi}-\dot{\phi}) \\ \times c_{\delta\Psi} \! \! \int_{\eta_0}^{\eta_r} \! d\eta \ j_{\ell} \, (k[\eta-\eta_0]) \, a(\eta) \, b_{\rm{lin}}(\nu) \, \bar{j}(\nu,\eta) \, \tilde{\delta}(k,\eta)
\end{multline}
where $\Delta_{\mathcal{R}}^2$ comes from the primordial curvature power spectrum $P_{\mathcal{R}} \equiv 2\pi^2\Delta_{\mathcal{R}}^2/k^3$~;
$j_{\ell}(\cdot)$ are the spherical Bessel functions, while $\tilde{\delta}$, $\phi$ and $\psi$ are the time-dependent
parts of (respectively) the matter density contrast $\delta$, and the two Newtonian gravitational potentials $\Phi$ and $\Psi$. The two coefficients $c_{\Psi\Phi}$ and $c_{\delta\Psi}$ give the relations between $\delta$, $\Phi$ and $\Psi$ for adiabatic initial conditions~: $c_{\delta\Psi} \equiv {\delta}/{\Psi}=-{3}/{2}$, $c_{\Psi\Phi} \equiv {\Psi}/{\Phi} = -(1 + ({2}/{5}) R_\nu)$,
where $R_\nu \equiv \rho_\nu/(\rho_\nu+\rho_\gamma)$, with $\rho_\nu$ and $\rho_\gamma$ respectively the energy densities in relativistic neutrinos and photons.


\section{Signal-to-Noise analysis}
\label{sec:sn}

To compute the aforementionned spectra, we adapted for our analysis a code named CROSS\_CMBFAST \cite{CCMBF}~: for a given cosmology and emissivity function $\bar{j}(\nu,z)$,
our code calculates the $C_{\ell}^{\rm{cr}}$ from Eq.~(\ref{crosscl}) and at the same time the predicted power spectrum of the CIB fluctuations described by Eq.~(\ref{dtcib}).

\subsection{Ideal case}
\label{sec:perf_sn}

We now investigate the detection level of the ISW effect using CMB-CIB cross-correlation by performing a signal-to-noise ratio analysis in the context of several past (IRAS) and present (\textit{Herschel} SPIRE, \textit{Planck}) experiments and their associated frequencies. Using our computed power spectra, we can write for each given frequency $\nu$ the total signal-to-noise ratio of the ISW detection as :
\beq{sn1}
\left[\frac{S}{N}\right]^2 \! \! \! \! (\nu) = \sum_{\ell=2}^{\ell_{\rm{max}}} (2\ell+1)\frac{[C_{\ell}^{\rm{cr}}(\nu)]^2}{[C_{\ell}^{\rm{cr}}(\nu)]^2+C_{\ell}^{\rm{CIB}}(\nu) \times C_{\ell}^{\rm{CMB}}}
\eeq
where the total (or cumulative) signal-to-noise is summed over multipoles between $\ell=2$ and $\ell_{\rm{max}} \leqslant 100$ where the signal has its major contribution.

We first consider the ideal situation where the CIB and CMB maps used for cross-correlation are noiseless and cover the whole sky ; with these ideal assumptions, we obtain high levels of detection for the CIB-CMB correlation which reach $\sim6$\,-\,$7\sigma$, with the largest contribution to the SNR coming from multipoles lower than $\simeq50$. Our results for the \textit{Planck} instrument are shown on Table~\ref{SNRtab}~; in light of these, the optimal frequency for ISW detection appears to be around 353~GHz with a maximum SNR reaching 7$\sigma$. However in pratice, the CIB extraction at this frequency might prove challenging since the CMB becomes dominant here, and increasingly so as we go down in frequency. Therefore the possible residuals in the extracted CIB map have to be accounted for, and other sources of noise as well, which we do next.


\subsection{More realistic SNR}
\label{sec:real_noise}

We now carry a more realistic study by including several possible sources of contamination~: first, the strong emission of the Milky Way masks the CIB signal over the galactic plane, reducing the usable fraction of the sky ($f_{\rm{sky}}$) by at least $\sim25\%$, and consequently reducing the SNR by a factor $\sqrt{f_{\rm{sky}}}$. Furthermore, the rest of the sky is also quite polluted by dust foregrounds, which will have to be removed from our CIB maps although some residuals might remain. There may even be a significant CMB residual in this map, due to an imperfect separation of components. Consequently, we assess the impact of these contaminants in our study, by adding a noise term $N_{\ell}^{\rm{CIB}}$ to the CIB power spectrum in our SNR. We break this noise component into several independant parts~:
\vspace{-0.25cm}
\begin{equation*}
N_{\ell}^{\rm{CIB}}(\nu) = R_{\ell}^{\rm{CMB}}(\nu) + R_{\ell}^{\rm{fore.}}(\nu) + N_{\ell}^{\rm{instr.}}(\nu) + N_{\ell}^{\rm{correl.}}(\nu)
\end{equation*}
where these four different terms represent the power spectra of the CMB residual, the galactic foregrounds residuals, the instrumental noise and finally the noise due to correlation between residuals and the CIB (which appears when autocorrelating the final CIB map). The instrumental noise is fixed by the characteristics of the considered experiment. We then choose three parameters to govern the other three contributions~: $f_{\rm{sky}}$, the fraction $f_{\rm{CMB}}$ of the total CMB map present as a residual in the CIB map, and $\mathcal{A}_{\rm{fore.}}$ the amplitude of the foreground residuals spectrum relative to the CIB spectrum (at $\ell=10$). The next step would be to explore this 3D parameter space at each frequency and compute the SNR at each point. Considering the very large number of possible combinations of parameters, it would not be practical to display the complete results of this exploration here. Therefore we first choose to fix $f_{\rm{sky}}$ to two values of interest~: $f_{\rm{sky}}=0.75$, corresponding to an optimistic case where only the galactic plane is discarded~; $f_{\rm{sky}}=0.15$, a pessimistic estimate of the area of the sky where the current data allows for an efficient CIB extraction. 
For our other two parameters we limit ourselves to reasonable values, with $f_{\rm{CMB}} \in [0,0.1]$ and $\mathcal{A}_{\rm{fore.}} \in [0,10]$. We also explore the combination of cross spectra at different frequencies, which allows us to increase the total signal-to-noise ratio of the ISW detection by combining their constraints, although this will be limited by the possible intrinsic correlations between the different CIB frequencies.

After a thorough exploration of these parameters, we show a part of our results in Table~\ref{SNRtab}) and we can draw a few conclusions~: First the gain brought by the joint correlation is small, which can be attributed to the high correlation (therefore redundancy) between the CIB at its different observed frequencies. When foucusing on the 353~GHz results (the best SNR in the ideal case), the influence of the CMB is clearly visible, quickly reducing the SNR as its residual level increases, which is even more pronounced at 217~GHz, due to the proximity to the maximum of the SED of the CMB. The presence of instrumental noise becomes significant at the two lowest frequencies (217 and 353~GHz), again reducing their value in the cross-correlation. As expected the galactic foreground residuals also decrease the SNR, though their influence is roughly the same at all frequencies as they are defined relatively to the CIB spectrum. Lastly, the biggest influence comes from the fraction of the sky through the $f_{\rm{sky}}$ parameter, as the total SNR scales as $\sqrt{f_{\rm{sky}}}$. This makes it a crucial requirement for future applications to have the largest possible coverage to minimize this effect.

\begin{table}
\small
\caption{Total signal-to-noise ratio of the CIB-CMB cross-correlation for four of the CIB frequencies of \textit{Planck} HFI and three different noise scenarii.}
\begin{center}
\begin{tabular}{ccccc}
\br
\textbf{Frequency (}GHz\textbf{)} & \textbf{857} & \textbf{545} & \textbf{353} & \textbf{217} \\
\textbf{Wavelength ($\mu m$)} & \textbf{350} & \textbf{550} & \textbf{850} & \textbf{1380} \\
\hline
\textbf{Perfect Single SNR} & 6.26 & 6.83 & 6.98 & 6.95 \\
\textbf{Joint SNR} & \multicolumn{4}{c}{7.12} \\
\hline
\textbf{Realistic SNR 1 ($f_{\rm{sky}}=0.75$,} & 5.36 & 5.73 & 5.39 & 3.56 \\
$f_{\rm{CMB}}=0.01$, $\mathcal{A}_{\rm{fore.}}=0.01$) & \ & \ & \ & \ \\
\textbf{Joint SNR} & \multicolumn{4}{c}{5.88} \\
\hline
\textbf{Realistic SNR 2 ($f_{\rm{sky}}=0.15$,} & 2.40 & 2.56 & 2.41 & 1.59 \\
$f_{\rm{CMB}}=0.01$, $\mathcal{A}_{\rm{fore.}}=0.01$) & \ & \ & \ & \ \\
\textbf{Joint SNR} & \multicolumn{4}{c}{2.63} \\
\br
\end{tabular}
\end{center}
\label{SNRtab}
\vspace{-0.8cm} 
\end{table}


\section{Conclusions}
\label{sec:conclusions}

We investigated the cross-correlation between the cosmological infrared and microwave backgrounds through the ISW effect, and studied its detectability under various observational situations. Using an advanced SNR analysis which included the main sources of noise both instrumental and astrophysical, and all their possible correlations, we pointed out the most promising frequency in the ideal case of noiseless full-sky maps (353~GHz) with an expected significance as high as $7\sigma$ for the cross-correlation signal. The same frequency turned out to be less optimal with more realistic assumptions about sky coverage and possible sources of noise (here CMB, dust residuals and instrumental noise). In this case, higher frequencies such as \textit{Planck} HFI's 545 and 857 GHz are favored, with an expected significance ranging from 2 to more than 5$\sigma$ depending on the frequency, the levels of noise and the fraction of the sky available for analysis. We also found that a joint cross-correlation using all available frequencies is of minor interest. Nevertheless, our best results for $f_{\rm{sky}}=0.75$ are higher than the significances of all current CMB-galaxies cross-correlation, with $\sigma>5$, although a less optimistic estimate for the sky coverage quickly reduces our signal-to-noise ratios. This stresses once again the requirement of good component separation techniques and foreground removals for future applications, in order to have the largest fraction of common clean sky $f_{\rm{sky}}$ possible. We are currently applying our method and framework to the actual \textit{Planck} data, and the results of this study will be published with the next public release of \textit{Planck} papers and data at the beginning of 2013.

\vspace{-0.15cm}

\section*{References}

\bibliographystyle{iopart-num}
\bibliography{article}

\end{document}